\documentclass[sigconf, preprint]{acmart}

\usepackage{todonotes}
\usepackage{makecell}
\usepackage{bm}

\newcommand{\methodname}{TwERC}

\AtBeginDocument{%
  \providecommand\BibTeX{{%
    \normalfont B\kern-0.5em{\scshape i\kern-0.25em b}\kern-0.8em\TeX}}}

\setcopyright{acmcopyright}
\copyrightyear{2023}
\acmYear{2023}
\acmDOI{XXXXXXX.XXXXXXX}

\acmConference[Conference acronym 'KDD]{Make sure to enter the correct
  conference title from your rights confirmation email}{August 06--08,
  2023}{Long Beach, CA}
\acmPrice{15.00}
\acmISBN{978-1-4503-XXXX-X/18/06}

\begin{document}

\title{\methodname: High Performance Ensembled Candidate Generation for Ads Recommendation at Twitter}



\author{Vanessa Cai}
\authornote{Both authors contributed equally to this research.}
\affiliation{%
  \institution{Twitter}
  San Francisco, CA, \country{USA}
}

\author{Pradeep Prabakar}
\authornotemark[1]
\affiliation{%
  \institution{Twitter}
  San Francisco, CA, \country{USA}
}

\author{Manuel Serrano Rebuelta}
\affiliation{%
  \institution{Twitter}
  New York City, NY, \country{USA}
}

\author{Lucas Rosen}
\affiliation{%
  \institution{Twitter}
  New York, NY, \country{USA}
}

\author{Federico Monti}
\affiliation{%
  \institution{Twitter Cortex}
  London, \country{UK}
}

\author{Katarzyna Janocha}
\affiliation{%
  \institution{Twitter Cortex}
  London, \country{UK}
}

\author{Tomo Lazovich}
\affiliation{%
  \institution{Twitter Cortex}
  Boston, MA, \country{USA}
}

\author{Jeetu Raj}
\affiliation{%
  \institution{Twitter}
  Seattle, WA, \country{USA}
}

\author{Yedendra Shrinivasan}
\affiliation{%
  \institution{Twitter Cortex}
  New York City, NY, \country{USA}
}

\author{Hao Li}
\authornote{Equal Contributions}
\affiliation{%
  \institution{Twitter}
  Seattle, WA, \country{USA}
}

\author{Thomas Markovich}
\authornotemark[2]
\authornote{Corresponding author: thomasmarkovich@gmail.com}
\affiliation{%
  \institution{Twitter Cortex}
  Boston, MA, \country{USA}
}

\renewcommand{\shortauthors}{Cai and Prabakar, et al.}

\begin{abstract}
Recommendation systems are a core feature of social media companies with their uses including recommending organic and promoted contents. Many modern recommendation systems are split into multiple stages - candidate generation and heavy ranking - to balance computational cost against recommendation quality. We focus on the candidate generation phase of a large-scale ads recommendation problem in this paper, and present a machine learning first heterogeneous re-architecture of this stage which we term \methodname. We show that a system that combines a real-time light ranker with sourcing strategies capable of capturing additional information provides validated gains. We present two strategies. The first strategy uses a notion of similarity in the interaction graph, while the second strategy caches previous scores from the ranking stage. The graph based strategy achieves a 4.08\% revenue gain and the rankscore based strategy achieves a 1.38\% gain. These two strategies have biases that complement both the light ranker and one another. Finally, we describe a set of metrics that we believe are valuable as a means of understanding the complex product trade offs inherent in industrial candidate generation systems.
\end{abstract}

\ccsdesc[500]{Information systems~Recommender systems}
\ccsdesc[500]{Information systems~Personalization}
\ccsdesc[500]{Information systems~Nearest-neighbor search}

\keywords{candidate generation, information retrieval, recommendation systems}

\maketitle

\section{Introduction}

Twitter is an online social networking platform that allows hundreds of millions of users to read, write, and share short messages called tweets. These messages can include text, images, video, and links. It generates revenue through advertising, which can sometimes negatively impact user experience. However, this negative impact can be mitigated by targeting ads to users such that they are relevant and timely.

This task of identifying, or recommending, items to a user is commonly solved through the use of recommendation systems. Modern advertising systems are built around recommendation systems, which seek to recommend an advertisement to a user given everything the system knows about the user, similar users, the advertisement, and advertisements like it~\cite{kang2020tree, broder2008computational}. Recommendation systems commonly make recommendations using approaches either based on collaborative filtering~\cite{su2009survey, he2017neural, konstan1997grouplens, herlocker2000explaining}, or content filter~\cite{balabanovic1997fab, pazzani2007content, basilico2004joint, lee2002neural}. Collaborative filtering uses the past behavior of similar users to predict the preferences of a targeted user. Content filtering, on the other hand, uses the characteristics of the items, such as their text, images, and other metadata, to predict the preferences of a targeted user. The ultimate goal of both strategies is a system that predicts a user's expected engagement rate. In the ads recommendation, the predicted engagement rate is usually combined with advertiser specified bid (willing to pay) to holistically achieve a desirable balance among users, advertisers, and the platform, so that users are delighted by the ads experience, advertisers are satisfied with the marketing return-on-investment, and the platform can monetize the traffic. The task of ranking all active ads in real-time has become computationally prohibitive as the ads inventories have grown and the model complexity has increased.

To address this issue, modern recommendation systems divide the task into two parts, candidate generation and ranking, as it allows for a more efficient use of computational resources within a tight latency budget, and can lead to better performance compared to a single-stage approach~\cite{liu2022neural}. Candidate generation systems aim to filter the space of all eligible items, which can number in the billions, down to a set of a few thousand likely relevant items for ranking. These methods are often evaluated by recall, which measures the proportion of relevant items that are successfully retrieved. The goal of candidate generation is to find the most relevant items among all the active items, with minimal computational cost. This process is closely related to the field of Information Retrieval (IR), which focuses on finding relevant information from a large collection of documents~\cite{chowdhury2010introduction, singhal2001modern, kobayashi2000information}. Common candidate generation technique include Boolean retrieval~\cite{lee1993evaluation, buell1981threshold, waller1979mathematical} and vector space models~\cite{wong1984vector, wong1987modeling, raghavan1986critical, ai2016analysis, billhardt2002context, barkan2016item2vec}.
Ranking models then take the relevant items that have been identified by the candidate generation process and use complex models (usually deep neural networks with millions of parameters) to predict the likelihood of engagement. Ranking systems are frequently evaluated by precision, which measures the proportion of retrieved items that are relevant. The goal of ranking models is to sort the relevant items by their predicted utilities (a combination of engagement likelihoods and advertiser specified bids in the ads recommendation problem). Though ranking models have received extensive study~\cite{anil2022factory, wang2017deep, wang2021dcn, naumov2019deep, cheng2016wide}, candidate generation systems have received less attention in both the academic literature and industrial applications~\cite{pinnersage}.

In this work, we focus on the candidate generation side of recommendation systems. Our research focus is on identifying techniques to efficiently generate high quality candidates for downstream ranking tasks in industrial settings. To address this question, we develop a system called Twitter Ensembled Retrieval of Candidates (\methodname), with a focus on ads applications. In section 2 we review the literature and related systems. In section 3, we present a rebuilt Twitter ads candidate generation stack system that includes a system for combating feedback loops in recommendation systems, a fully counterfactual candidate generation method, a graph similarity-based candidate generation method, and a blender that combines all of these sources together. These pieces combine together to form \methodname. We review both offline and online experiments  for \methodname in section 4, and show that this rebuilt system improves the efficiency and effectiveness of the candidate generation process by reducing the number of irrelevant advertisements that are shown to users, while at the same time increasing the number of relevant items that are shown to users. Finally, we present our conclusions in section 5.

\section{Related Work}
\textbf{Counterfactual Data Collection and Policy Evaluation in Recommendations:} Feedback loops and online/offline gaps in recommendation are significant problems in industrial recommender systems. The feedback loops ultimately inhibit our ability to perform counterfactual experiments. Counterfactual policy estimators are used to sidestep this issue. Genie employs an open box simulation engine with click calibration model to compute the KPI impact of any modification to the system, using the auction participants data (impression winners and losers)~\cite{Bayir2019}. Another approach involves changing the loss function to enable counterfactual learning-to-rank for both support vector machines~\cite{Joachims2017} and a more general class of classifiers including neural networks~\cite{Agarwal2019}. Our offline evaluation is similar with Genie's setup, which allows us to confidently perform offline experiments, but leverage it for candidate generation. To our knowledge, this is the first such use.

\textbf{Graph Machine Learning:}
Graph machine learning has received significant academic and industrial interest in recent years, with applications including everything from healthcare~\cite{li2020graph, wang2020recent} to finance~\cite{wang2021review, cheng2022financial, matsunaga2019exploring} and social media~\cite{twhin2022, fan2019graph, han2020graph}. The most common paradigm involves computing an embedding of the application relevant graph, and using those embeddings in downstream applications. One family of approaches to generate embeddings that was inspired by NLP methods is based on the embedding of metapaths~\cite{Deng2022, Naseem2021}. The different methods can be understood through the metapath sampling strategy and loss function. The earliest of these methods include the SkipGram~\cite{skipgram}, Word2Vec~\cite{word2vec}, and DeepWalk~\cite{Perozzi2014} methods, which construct meta-paths through depth first search. Methods like LINE~\cite{Tang2015} and PTE~\cite{Tang2015_2} instead perform a breadth first search. Finally, some methods like Node2Vec~\cite{Grover2016} and Metapath2Vec~\cite{Dong2017} perform a hybrid searching strategy while also allowing the construction of metapaths not present in the original graph connectivity. Graph neural networks such as GraphSage~\cite{hamilton2017inductive}, GraphSaint~\cite{zeng2019graphsaint}, and SIGN~\cite{frasca2020sign} among others have been developed to perform a nonlocal weighting of those random walks and have found extensive applications in recommendation systems~\cite{wu2022graph}. These methods largely cater to homogeneous graphs - those with a single edge type and a single vertex type -- with the exception of PTE and Metapath2Vec.

Inspired by the logical calculus of knowledge graphs, methods like TransE~\cite{bordes2013translating}, TransH~\cite{wang2014knowledge}, TransR~\cite{lin2015learning}, and TransD~\cite{ji2015knowledge} all seek to learn a set of entity embeddings and relational operators that map the head of an edge to the tail. This entire family is referred to as translation methods because they learn a parameterized relation dependent translation. Unlike the metapath based methods, translation methods were constructed to handle heterogeneous data. Additionally, several different frameworks have been developed to scale these translation methods to webscale graphs~\cite{lerer2019pytorch, zheng2020dgl, zhu2019graphvite}.

\textbf{Candidate Generation:} Graph embeddings have been used in recommendation systems, where they find a natural application as the proximity of a user to an item in the embedding space implies the likelihood that a user will enjoy the recommendation. Methods like TwHIN~\cite{twhin2022}, GraphSage, and Personalized entity recommendation~\cite{Yu2014} demonstrated that injecting graph embeddings into ranking models provided significant metric wins. Techniques like Binary Encoding~\cite{kang2019}, Pinnersage~\cite{pinnersage} and Spherical graph embeddings~\cite{Zhu2022}, or SimClusters~\cite{simclusters} have found that graph embeddings are particularly useful for improving the quality of candidate retrieval.

\begin{figure}[h]
\centering
\includegraphics[scale=0.3]{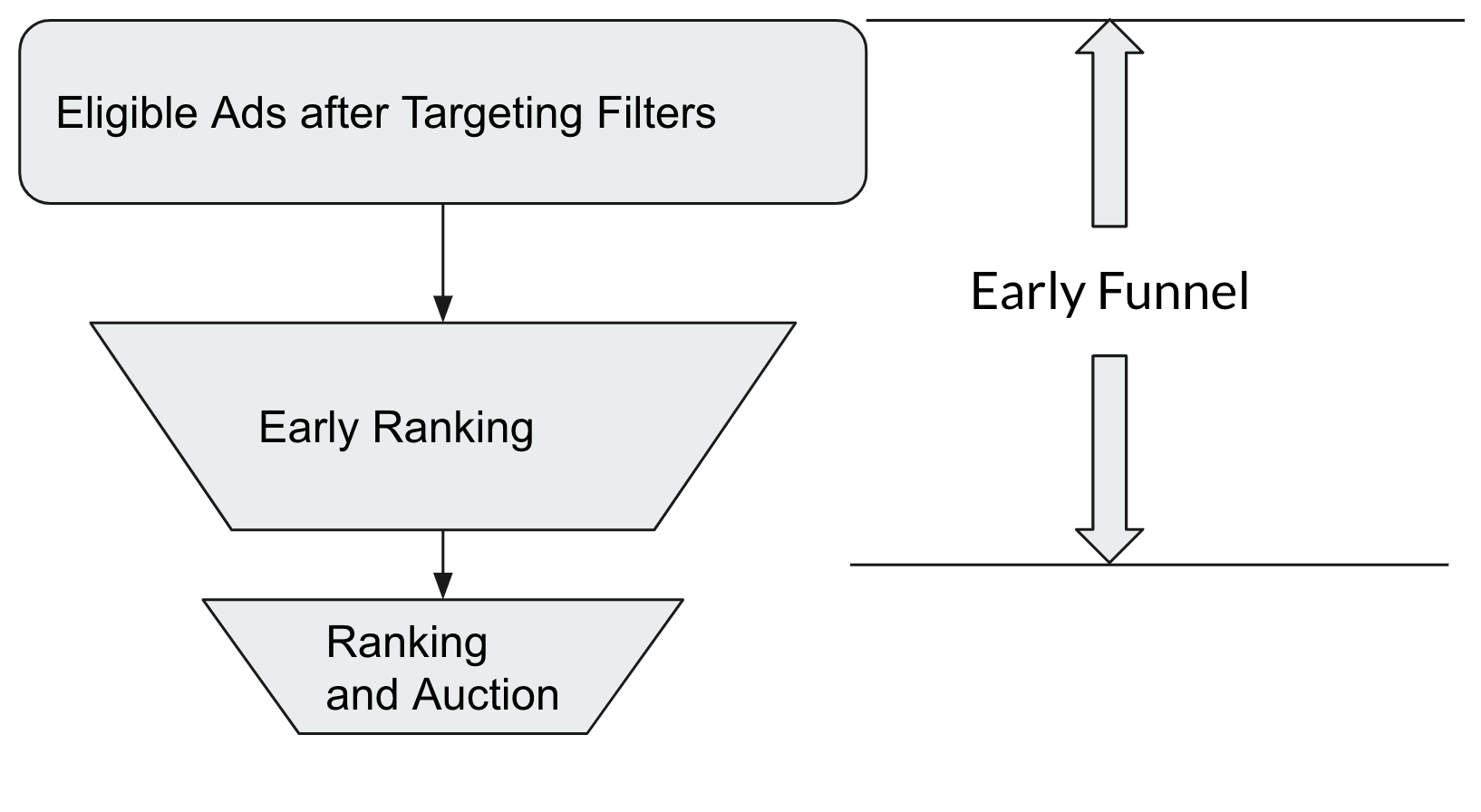}
\caption{Ads Serving Overview at Twitter Timeline Home}
\label{fig:ad-outline}
\end{figure}

\label{sec:theory}
After a user opens up or refreshes their home timeline, the page displays a combination of organic recommendations and advertisements. The organic content and advertisements are ranked concurrently, and mixed together at the end of the pipeline with each ad being inserted into the associated advertising slot. The recommendation system built for advertising recommendations contains three major stages as indicated in Figure~\ref{fig:ad-outline}: targeting and filtering; early ranking; and heavy ranking and auction. The final phase of recombining promoted and organic content into a single feed is managed by a separate system and out of scope of this work.

The targeting and filtering applies both hard and soft constraints to the active advertisements, to ensure that we only recommend ads to a user that the advertisers intended the user to see. Hard constraints can be thought of as \textit{and-clauses} and include age, geography, or language, and are required to all be satisfied. Soft constraints can be understood as \textit{or-clauses}, and include constraints such as follow-relationships or topic interests. The set of ads that pass this early filtering stage are then sent to a light ranker. This light ranker is a two-tower model that has been tuned for recall and optimized for fast inference~\cite{li2022inttower}. The topK advertisements are then sent to our heavy ranking system, which computes the calibrated probability of engagement for the ad given the optimization objective of that advertisement. In contrast to the light ranker, these heavy ranking models include many heavier-weight features such as richer content representations or graph based user representations~\cite{twhin2022}, and a much more complex architecture~\cite{wang2021dcn}. The heavy ranking models output a probability of engagement, pEng, that we then use to compute the rankscore for the second price auction. The rankscore is our best estimate of the overall value or utility of showing a particular ad, and is a function of the advertiser's bid and other factors like predicted engagement rates.

The above system is the product of years of development by many Twitter engineers. While we are always making improvements to the system, we have observed that making improvements to the input feature size and expressiveness of the light ranker is not generally feasible due to the tight latency budget. An alternative way forward to improve the early ranking stage is to perform what we call tail replacement. Tail replacement in this setting specifically means replacing the bottom M\% of the topK ads from the light ranker with a different strategy that is complementary and efficient. In this section we explore two separate, but complementary, approaches for tail replacement -- rankscore candidate generation and graph based candidate generation, and highlight practical considerations associated with putting these techniques into production. These two strategies comprise \methodname. We then discuss the process by which we combine candidates from all three different sources, and finally turn our attention to metrics and figures of merit.

\subsection{Rankscore Candidate Generation}

\textbf{Unconstrained Ad Serving}\label{sec:uas} The unconstrained ad serving system (UAS) is  a counterfactual data collection service that scores all the eligible ad candidates at the request level with a low sample rate. The selected request is duplicated and is then sent to a staging environment to get all ad candidates' bid and predicted engagement rate (positive and negative). In median, UAS scores approximately 50x more than the production scoring volume, granting the capability of observing the quality of ads that might have been filtered out by the early rankers, thus overcoming the infamous selection bias problem in large-scale online interactive/recommendation systems. With the logged bid and predicted engagement rate, we are able to calculate the $rankscore$ of all the ad candidates. This data allows us to bypass the common recommendation systems feedback loop, because we are able to rank, and scribe, all relevant ads for a user; rather than a filtered subset. This allows us to train the early ranking system on all data, rather than just its previous outputs. Additionally, we used these data to identify significant headroom (or regret of the current system) by optimizing the top$K$ candidates in the early funnel.

\textbf{Candidate Generation} The counterfactual data is not only used in measuring the regret and/or opportunity size, but also processed to generate user-level top$K$ ad candidates by expected value, through a process of downsampling requests and by removing invalid or incomplete requests. To be more specific, for each user $u$ and eligible ad $a$ we collect  $rankscore(u, a, t)$ at time $t$ from UAS with a 3-week lookback window, and calculate the time-aware user-ad quality score $q(u,a)$, 

\begin{equation}\label{eq:rs}
    q(u,a) =  \frac{\sum rankscore(u,a,t) \cdot e^{t - t_0}}{\sum e^{t - t_0}}
\end{equation}

The quality score is a weighted average at the 
$(u, a)$ level, with the weight being the elapsed time between the data collection time $t$ and the pipeline running time $t_0$. Effectively, this self-normalized score gives more weight to the fresher data points. Practically, we observe the time-aware weighting plays a significant role in adapting the volatile nature of the rankscore, and contributes significant incremental business impact on top of a simple average aggregation.  Then for each UAS covered user, we generate top$K$ ads with the highest $q(u,a)$ scores. We set up recurring jobs to generate these high value ad candidates every 3 hours and store them in a \emph{Manhattan} dataset (MH)~\cite{manhattan}, and the MH dataset is consumed by the ad mixer at serving time. 
Though all the UAS scored ads were eligible as of the data collection time, advertisers sometimes update their budget and targting criteria. To make sure we always fully respect advertisers' targeting criteria and their budget consumption status, we apply an online filter at the serving time too.

\begin{figure}[h]
\centering
\includegraphics[scale=0.3]{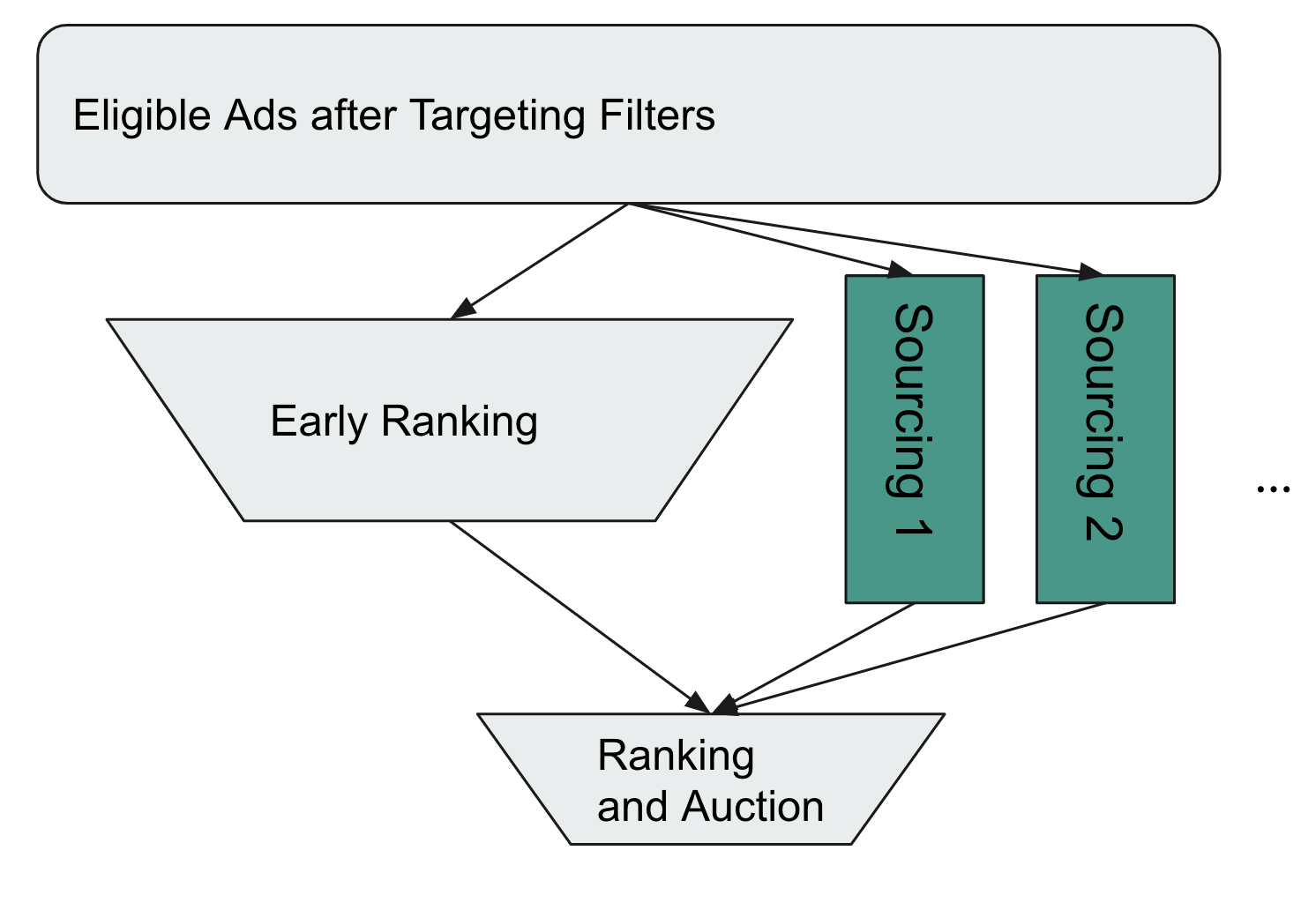}
\caption{New sourcing components (\textcolor{teal}{highlighted}) in the serving funnel.}
\end{figure}

\subsection{Graph Based Candidate Sourcing}\label{sec:graph_cg}

Due to the tight inference budget, the light ranker does not have enough parameters to represent the nuanced relationships that would normally be captured by a traditional collaborative filtering system. To address this deficiency, we choose to construct a candidate generation system that replaces up to M\% of the light ranker's tail with ads that have been identified as similar to ones similar users have engaged with before. This system fundamentally hinges on the concept of ``similar''. Inspired by previous work at Twitter~\cite{twhin2022} and elsewhere~\cite{pinnersage}, we choose to construct a heterogeneous graph embedding and use this as the core of our similarity search.

\textbf{Graph Embeddings} Retrieving items of interest to users is the central challenge with candidate generation. There are many ways to find signals that would indicate interest, or relevance. For the graph based candidate sourcing strategy, we choose to construct a directed multigraph from the engagements between users and the advertisements they have engaged with. This is done by processing the event level data into a heterogeneous engagement graph with different vertex types (e.g. user, advertisement, advertiser, application, video), and carefully designing the edge label definition. While we did not downsample data, we did clean data of all incomplete or otherwise invalid events. Following the recipe of TwHIN~\cite{twhin2022}, we then embed this graph in a lower dimensional space. We generated these TwHIN embeddings using the translation operator with the softmax loss function and dot-product distance function. We use a regularization parameter of $10^{-3}$, and generate 2000 negative samples for each positive one with 1000 generated using batch negative sampling and 1000 generated using uniform negative sampling. We tuned these hyperparameters through offline experiments, and selecting parameters that provided a satisfactory tradeoff in model performance and computational cost. See \cite{twhin2022} for more information. We interpret the distance between a user and an item as the likelihood that user would engage with that item. This strategy produces manifolds in which similar users and similar items are clustered together. This allows us generate candidates through a nearest neighbor search

\textbf{Embedding Update Cycles} The ads engagement graph is constructed by aggregating engagements over a fixed time window. The choice of time window naturally prohibits the representation of interactions as well as vertices which fall outside that window, which make it challenging to keep the graph and associated embeddings up to date.

We choose to address this issue through a warm-starting procedure. Because we have new vertices associated with new users or advertisements, we have to compute new embeddings. Starting the warm-start optimization process with randomly initialized embeddings for new entities would move the already converged vertices, thereby increasing the amount of noise in our embedding updates. For this reason, we freeze the embeddings of the old vertices while the embeddings of the new vertices are computed. One can view this as projecting the new vertices into the existing manifold. Because the old embeddings are frozen, we only need to include edges that include any of the new vertices. We term this set of new edges the $\Delta$ edges. This is where the ``tic'' update cycle stops. These tic embeddings are quick enough that we compute them every twelve hours.

The tic cycle clearly is incapable of capturing the relaxation of the entire graph due to the new interactions and new vertices. To do this, the toc cycle constructs an edge list by combining the $\Delta$ edge list with the complete edge list from the previous embedding. With this composite edge list in hand, we then unfreeze the old embeddings and let the entire graph relax until it converges.

At present, we perform a weekly toc cycle, and then tic cycles every 12 hours in between.

\textbf{Candidate Generation} Intuitively, we can retrieve a high-recall set of ad candidates by using a user’s TwHIN embedding to query its neighboring promoted tweets using an ANN index~\cite{athitsos2008nearest, johnson2019billion}. To efficiently retrieve these ads without being constrained by online latency requirements, we built a mapreduce pipeline that precomputes these in batch using the FAISS~\cite{johnson2019billion} library for our 130 million heaviest users. We configured FAISS to use the inverted file index + HNSW to improve speed while maintaining high recall. Additionally, we implemented a filtering step to remove ads that are ineligible for a particular user due to targeting criteria or insufficient budget, in order to avoid cache misses during the retrieval stage. We run this pipeline every six hours to both leverage the embedding refresh cycle and avoid storing stale or expired ads. 

\begin{figure}[h]
\centering
\includegraphics[scale=0.35]{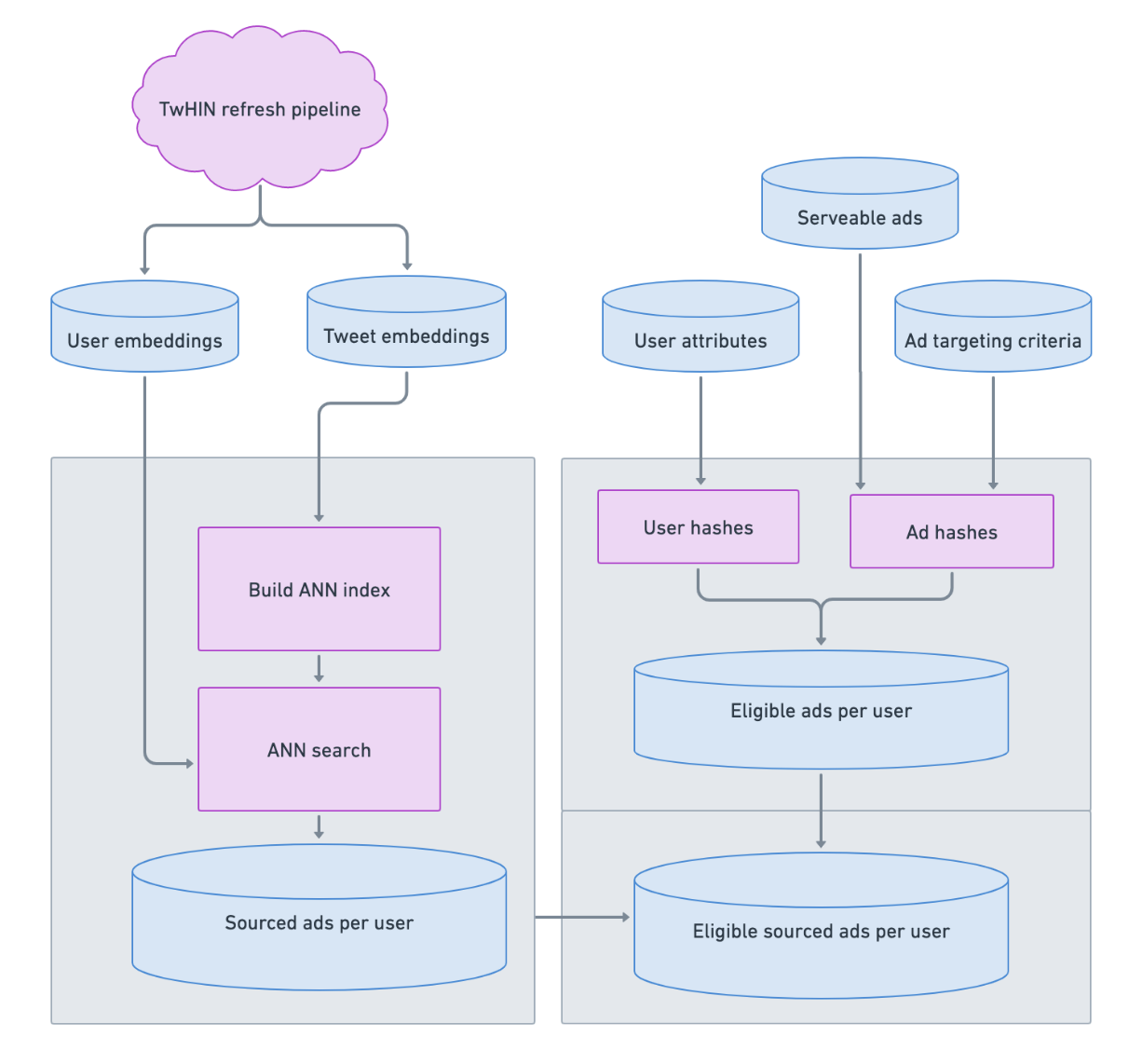}
\caption{Graph similarity based candidate generation pipeline.}
\end{figure}

\subsection{Blending}

With the initial promising online experiment results from both rankscore candidate generation and graph based candidate sourcing sourcing, we built a blender component in the ads serving early funnel, which dynamically allocate and effectively merging the sourced ads capacities from various sourcing strategies. We use a configuration file ($strategy-percentage$ pairs for each strategy) to specify the capacity allocation among the sourcing strategies, and use it in both offline batch data processing and online serving/merging logic. As an example, in the first blending experiment, we use a 20\%-20\% blending between the rankscore sourcing and graph based sourcing strategies, with the following key-value configuration  \{"rankscore": 0.2, "graph": 0.2\}. With the specified capacity allocation, we use them to generate a merged \emph{Manhattan} dataset~\cite{manhattan}, and fill in at most 20\% of the candidates with counterfactual-based and graph-based respectively for the full ranking and auction to further decide the final ads impressions (auction winners). We have experimented with a variety of different blending ratios, and we have empirically observed that a 20\%-20\% blending is optimal.

\subsection{Efficiency}

Both algorithms that we have proposed can be implemented as batch data pipelines, which feed a distributed key-value store that is utilized for online serving. Because the recommendations are run in batch, the impact to real-time serving is limited to the cost of key-value lookup, which is minimal. The computational resources required are significantly lower than the business impact they provide. Because these pipelines are run in batch, any failures will result in stale recommendations but will not cause serving failures. We have observed an improvement in product metrics if we run the pipelines more frequently, because they provide fresher candidates. We have experimented with a range of refresh-periods and selected the refresh cycle for each algorithm the represents a reasonable tradeoff between computational costs and product metrics. Finally, these pipelines do require maintenance but we have empirically observed that they are quite stable because they are built on top of a robust, distributed, data processing framework.

\subsection{Metrics}

The choice of metrics is both a deeply interesting question and critical to the success of any project because they will guide the overall progress of research and development. It is important to find offline metrics that correlate with our various online metrics and product concerns.

\textbf{Recall} The most obvious and easiest to define metric is recall, or hit-rate, defined as:
\begin{equation}
    R = \frac{\textrm{TP}}{ \textrm{TP} + \textrm{FN} },
\end{equation}
where TP is the number of true positives, or hits, and FN is the number of false negatives. We choose to define a hit as an engagement that was correctly identified in the test set. This provides a method to characterize the ability to predict user engagements. While this is valuable, optimizing for pure recall can be sub-optimal because it neglects the role of the auction and the true positive signals can be sparse and has the potential to underweight brand ads.

\textbf{Auction Recall} To address these issues, we introduce the concept of an auction recall. Specifically, we consider true positives to be those ads that win the auction for their slot. Optimizing for auction recall introduces multiple possible issues, including constructing a system that learns the biases and pathological behaviours of the ranking stack that is downstream. While this can lead to short term metric gains both online and offline, it can easily lead to long-term product decay. As a result, we typically examine both auction recall and engagement recall metrics.

\textbf{Rankscore NCG} Rankscore normalized cumulative gain (NCG) compares a given candidate generation algorithm to a hypothetical algorithm that always selects the top ads by \emph{rankscore}. The metric that is computed is the ratio of the sum of the rankscores of the ads selected by the algorithm and the potential rankscore sum that would be found by the hypothetical algorithm. It is defined as:
\begin{equation}
    rNCG_m = \frac{\sum_{i=1}^{n}\sum_{j=1}^{min(m,|C_i|)}C_{i,(|C_i|-j+1)}}{\sum_{i=1}^{n}\sum_{j=1}^{min(m,|R_i|)}R_{i,(|R_i|-j+1)}},
\end{equation}
where $R_i$ is the set of rankscores of all eligible ads for the ith request after targeting filters are applied, $C_i$ is a the subset of ad candidates selected by the candidate generation algorithm from $R_i$, $n$ is the number of ad requests, and $m$ is set to match the number of ads that would typically make it to auction. Lower rankscore NCG ratios indicate a larger headroom for showing more highly ranked ads by making improvements to candidate generation.

\textbf{Inequality} Finally, we track the top 1 percent share (T1PS) of the advertisers to evaluate whether our changes are making the overall ads ecosystem more fair. This is important to improving the advertising experience on the Twitter platform because it improves the experience of small and medium sized businesses, which is a long standing product goal for all online advertising platforms. The T1PS coefficient is defined as:
\begin{equation}
 T1PS = \frac{\sum_{S_i \geq S_{(0.99|S|)}} S_i}{\sum_i S_i},
\end{equation}
where $S_i$ is the number of served ads among all users from advertiser $i$ over a given period. Thus the condition, $S_i \geq S_{(0.99n)}$ restricts to advertisers in the 99th percentile or top 1 percent of advertisers with regards to served ads. 

\textbf{Ads Value} It is generally difficult to estimate the average value an advertiser realizes from running an ad campaign on the platform. While we cannot estimate that exact value because we do not have access to the per-advertiser exchange rate for engagements, we can use the average cost per conversion as a natural proxy for that value. Using this intuition, we derive a measure of proxy ads value:
\begin{equation}
    \textrm{AdsValue}(j) = \sum_i \frac{R_i}{C_i} C_{ij}
\end{equation}
where $j$ indicates the experiment bucket, $i$ indicates the campaign index, $R_i$ indicates the revenue for that campaign, $C_i$ indicates the number of conversions for that campaign, and $C_{ij}$ is the number of conversions for the $i^{th}$ campaign in the $j^{th}$ experiment bucket.

An increase in ads value corresponds to improvements to our ads ecosystem on the demand side. Tracking the ads value allows us to understand if changes to our recommendation ecosystem are improving the overall long-term value that we provide to our advertisers in an a/b test setting.

\textbf{Utility} Utility is a derived metric that provides a way to estimate the statistical performance of our end to end system. To understand utility, we first have to understand rankscore. Because Twitter is a second price auction, we need to provide our best estimate for the overall value of a particular ad. This is done by combining factors such as the probability of engagement (pEng), probability of negative engagement (pNeg), the advertisers bid. The utility is simply the rankscore with observed values used in place of the probabilities. An increase in utility corresponds to an increase in the quality of our predictions.

\section{Experiments}
Initial investigations found that the ads serving early funnel has significant headroom in the rankscore sum ratio, an ads early funnel metrics indicating the amount of regret we experienced due to filtering. An additional set of experiments found that truncating the tail M\% of the early ranker's top$K$ candidates that are sent to full ranking resulted in no observed decrease to net revenue. This indicated an opportunity to improve the filtering phase by experiment with additional candidate generation strategies. We built \methodname, which provides relevant candidates with controlled computational costs to meet this need. To do this, we explored four different strategies that augmented the early ranker. Two of these strategies are deployed in production.

We ran a range of offline and online experiments to evaluate the impact of the two separate sourcing strategies that were constructed. While the experiments presented below are run with each sourcing strategy in isolation, we later on observe that the impact of the two sourcing to be additive due to the construction of the blender.

All online experiments were run using a 2\% request sampling strategy, where 2\% of the traffic received the treatment. We ran each experiment for a minimum of seven days to account for statistical power and weekly user behavioral trends, and tracked a variety of different metrics to assess improvements to key product metrics. The results presented below are all statistically significant under a Benjamini \& Yekutieli correction \cite{benjamini2001} for multiple comparisons.

In the experiments presented below, we consider the current production early ranker to be our baseline. In the control arm, the early ranker ranks and selects the topK candidates at the advertisement level. In a nutshell, the early ranker is a two-tower neural network powered light pClick model, trained on the actual ads impression log. The trained pClick model outputs are later combined with ad candidates’ bid to rank the eligible ads candidates. At the serving time, ads embeddings are pre-computed and cached in memory, and the user embedding is computed on the fly. Other baselines such as Pinnersage~\cite{pinnersage} were not considered due to their computational cost.

Finally, for reasons of confidentiality, we are not able to specify the objective names below. In this work, an objective is an advertising goal that an advertiser might want to optimize for. Common objectives on advertising platforms such as Twitter include Click, Video View, or App Install, among others. While we have anonymized the objectives below, we have taken care to ensure that each table presents the same objective with the same id.

\subsection{Rankscore Candidate Sourcing}
Using the unconstrained ad serving system (see Section~\ref{sec:uas}), a counterfactual dataset containing full rankscores of ads, we store high full rankscore ads at the user level, and complement the top$K$ candidate decision previously generated solely by the early ranking model. In practice, we use a trailing 21-day window of the UAS data to generate the user level sourced ads dataset through a time-aware weighted average aggregation which give more weights to more recent rankscore for the user. It is powered by a scheduled batch job that refreshes every 3 hours. At the serving time, we replace at most M\% of the early ranker's tail top$K$ ads by these offline sourced ads, and send the merged candidate ads set to the full ranking.

We prepare the sourcedAds at the creative-user level. For the lineitems that are associated with sourcedAds, we allow them to bypass the lineitem ranker, but respect all restrictive targeting clauses (also known as targeting filters).

We first tested this strategy using our offline simulator, where we observe improvement, as high as 20\% in auction recall across different advertising objectives, as shown in Table \ref{tab: rs_offline}. 

\begin{table}[H]
\centering
\begin{tabular} {lc}
\Xhline{1.5pt}
Objective & AR $\Delta$ \\
\hline
Obj 1 & 11.1\% \\ 
Obj 2 & 2.3\% \\ 
Obj 3 & 4.7\% \\ 
Obj 4 & 20.1\% \\
Obj 5 & 15.9\% \\ 
\Xhline{1.5pt}
\end{tabular}
\caption{Results from the rankscore based candidate generation offline experiment. AR $\Delta$ is the auction recall delta between the production baseline and the experimental system.}
\label{tab: rs_offline}
\end{table}

Based on these offline experiments, we ran a range of online experiments with a range of different aggregation approaches. In the final iteration that we shipped to production (Table \ref{tab:rs_online}, we observed an 1.38\% increase in net revenue, a 4.71\% increase in utility per mille impressions, and a 1.05\% increase in ads value - all statistically significant with adjusted p-values. 

\begin{table}[H]
\centering
\begin{tabular} {lccc}
\Xhline{1.5pt}
Metric & Revenue & Utility & Ads Value \\
\hline
$\Delta \%$ & 1.38\% & 4.71 \% & 1.05 \% \\
\Xhline{1.5pt}
\end{tabular}
\caption{Results from the rankscore candidate sourcing online experiment}
\label{tab:rs_online}
\end{table}

\subsection{Graph Based Candidate Sourcing}

Previous work using graph embeddings in an offline setting showed promising results on a variety of ranking and candidate generation tasks. Inspired by this work, we looked to explore using these embeddings to generate candidate ads. We began by simulating auction recall after replacing the bottom M\% of the tail of the early ranker with nearest neighbor candidates as outlined in section~\ref{sec:graph_cg}. In the offline analysis, we found an auction recall improvement  across a variety of different objectives.

\begin{table}[H]
\centering
\begin{tabular} {lc}
\Xhline{1.5pt}
Objective & AR $\Delta$ \\
\hline
Obj 1 & 1.1\% \\ 
Obj 2 & 5.4\% \\ 
Obj 3 & 2.3\% \\ 
Obj 4 & 4.5\% \\
Obj 5 & 2.0\% \\ 
\Xhline{1.5pt}
\end{tabular}
\caption{Results from the graph based candidate generation offline experiment. AR $\Delta$ is the auction recall delta between the production baseline and the experimental system.}
\end{table}

Motivated by the offline experiments, we hypothesized that by adding graph signals, we would be able to better capture relational information about ads and users that can increase the quality of retrieved ads. To test this hypothesis, we constructed an online experiment where we precompute sourced ads by finding approximate nearest ad neighbors in the graph embedding space for one third of our highly active users offline. These candidates complement the top$K$ candidate decision previously generated solely by the early ranking model. In this experiment, we replaced at most M\% of the tail top-ranked ads by the early ranker by these offline sourced ads, and sent the merged candidate ads set to the full ranking. We made no change to the downstream full ranking and auction.

\begin{table}[H]
\centering
\begin{tabular} {lccc}
\Xhline{1.5pt}
Metric & Revenue & Utility & Ads Value \\
\hline
$\Delta \%$ & 4.08\% & 6.42 \% & 1.26\% \\
\Xhline{1.5pt}
\end{tabular}
\caption{Results from the graph based candidate sourcing online experiment}
\end{table}

We observed a 4.08\% increase in net revenue, a 6.42\% increase in utility per mille impression, and a 1.26\% increase in ads value - all statistically significant with adjusted p-values. We additionally analyzed this experiment in the context of advertiser inequality, and found a T1PS reduction of 1.2\% which was a significant effect. The control bucket result is also consistent with the historical T1PS. When we break down by advertiser type, we see that the small and medium businesses and mid-market size advertisers experienced the largest decreases in inequality, while direct sale and reseller accounts also experienced moderate decreases. This indicates that our experiment helped the advertisers who were experiencing the largest levels of inequality previously. This experiment is a very concrete example of a case where we can simultaneously increase net revenue and decrease advertiser inequality -- two objectives often thought to be in conflict.

\subsection{Time Dependent Embeddings}
The user embeddings that are used as query vectors generally converge to the average of all of the advertisements that the user engages with. While this provides a reliable way to predict a user's most common commercial interests, it misses interests that may have a temporal nature to them. For example, advertisements for infant goods may become less relevant over time as a parent's child ages. To address this issue, we followed the example of the rankscore temporal discounting in equation~\eqref{eq:rs} and computed time-decayed user embeddings as
\begin{equation}
    \vec{u}^{user}_{i, t} = \frac{
    \sum_{j \in \mathcal{N}_i} e^{\lambda \left( t_j - t_0 \right)} \; \vec{v}^{tweet}_{j} }{ \sum_{j \in \mathcal{N}_i} e^{\lambda \left( t - t_0 \right)} },
\end{equation}
where $|\mathcal{N}|$ is the set of the last-N engagements, $t_0$ is the time the job runs, $t_j$ is the time of the time of the $j^{th}$ interaction, and $\lambda$ is the decay constant. We implemented this aggregation strategy using a cloud SQL framework. Using this, we ran tail-replacement offline experiments using the same simulator that was used for the other candidate generation experiments.

\begin{table}[H]
\centering
\begin{tabular} {lc}
\Xhline{1.5pt}
Objective & AR $\Delta$ \\
\hline
Obj 1 & -1.0\% \\ 
Obj 2 & 7.9\% \\ 
Obj 3 & 10.2\% \\ 
Obj 4 & 7.2\% \\
Obj 5 & 9.9\% \\ 
\Xhline{1.5pt}
\end{tabular}
\caption{Results from the time-decayed graph based candidate generation offline experiment. AR $\Delta$ is the auction recall delta between the production baseline and the experimental system.}
\end{table}

We additionally explored the overlap between candidates using the time-dependent and time-independent. For users with a long history if advertising engagements, we observed little overlap between the two candidate generation strategies. For users with little to no history, however, the overlap increased significantly. This is not unexpected, because the time-dependent vector should be close to the time-independent vector. Finally, we performed a qualitative analysis of the candidates that are generated using these strategies. For users for whom we know their interests, we find this strategy captures emerging interests in some instances, as desired. We are encouraged by these results, as well as the low computational cost to generate these embeddings.

\subsection{Embedding Induction}

As illustrated in the previous section, using the candidates returned by an ANN-search in the graph embedding space increases the quality of the tweets returned by the early ranker and in turn, the revenue generated by the platform. However, not all users can be served via ANN-search, as users that have never interacted with an ad or an advertiser do not appear in the  engagement graph used to train graph embeddings, and thus do not enjoy a representation describing their interests in terms of promoted tweets (a common situation for new and light users for instance). To enhance the quality of recommended ads to a broader pool of users, we resorted to Graph Learning techniques to extend the coverage of our embeddings. In particular, we implemented an efficient formulation of the Feature Propagation approach described in \cite{rossi2022unreasonable} that propagates information  over the arcs of the follow graph. We decided to limit feature propagation to just one hop and infer graph embeddings for missing users $U_M$ as:
\begin{equation}
\label{eq:fp}
    \bm{x}_i = \frac{1}{|N^+_{S, T}(i)|} \sum_{j \in N^+_{S, T}(i)} \bm{x}_j \;\;\;\; \forall i \in U_M
\end{equation}

where $\bm{x} \in \mathbb{R}^d$ and $N^+_{S, T}(i)$ is a randomly sampled  set of 100 followings of $i$ that do have an embedding (here sampling is introduced to upper-bound the resources required to process each user). The main intuition  is to use the interests of the followings of a target user as a proxy for the interests of the user themselves. With this view in mind, the solution illustrated in (\ref{eq:fp}) can be understood as a form of collaborative filtering where the notion of user-user similarity is defined by follow connections and the averaging mechanism implements a voting system. Empirically, we observed this approach to perform particularly well in offline experiments and be extremely efficient at the same time as equation (\ref{eq:fp}) can be easily implemented with a simple SQL query, which in turn allows to exploit scalable frameworks such as  BigQuery for experimentation. 

Table \ref{tab:emb_ind_offline_users_with_twhin} and \ref{tab:emb_ind_offline_users_no_twhin} respectively show the results we obtained for users that have and do not have (before diffusion) graph embeddings at experimentation time. Only tweets returned by the ANN-search and that can be recommended to a user as specified by the advertiser's targeting criteria have been considered as possible candidates for a user. A tweet is considered as relevant for a target user if the user has interacted with this in the past. 

Diffused embeddings on the follow graph achieve very good performance in this offline setting, obtaining a HR@K equal to $\sim 79\%$ of the one showed by the true graph embeddings and maintaining similar performance also on users that do not have a graph embedding at all. At the time of writing, online evaluation of the diffused embeddings is ongoing.

Qualitatively, we observed diffused embeddings to produce good quality results on a few users for which we know their interests, although they might not always capture peculiarities of the users themselves. For instance, users based in a country but that predominantly follow people based abroad tend to get recommended tweets that are consistent with their interests but that are not necessarily targeted to residents of their specific country. This particular phenomenon could be attenuated introducing a weighting scheme on the followings of a target user that takes into account similarities in demographic or potentially engagement. We leave an exploration of this weighting scheme to future work.

\begin{table}[H]
\centering
\begin{tabular} {lcccc}
\Xhline{1.5pt}
Method & HR@300 \\
\hline
Random ordering & 6.5\% \\
Diffused graph embeddings & 34.38\% \\
Graph embeddings & 43.53\% \\
Collaborative Early Filtering & 49.52\% \\
\Xhline{1.5pt}
\end{tabular}
\caption{Offline results on diffused embeddings for users that do have a graph embedding (i.e. users that appear in the ad engagement graph).}
\label{tab:emb_ind_offline_users_with_twhin}
\end{table}

\begin{table}[H]
\centering
\begin{tabular} {lcccc}
\Xhline{1.5pt}
Method & HR@300 \\
\hline
Random ordering & 10.11\% \\
Diffused graph embeddings & 42.07\% \\
Collaborative Early Filtering & 55.61\% \\
\Xhline{1.5pt}
\end{tabular}
\caption{Offline results on diffused embeddings for users that have no graph embeddings before diffusion (i.e. users that do not appear in the ad engagement graph).}
\label{tab:emb_ind_offline_users_no_twhin}
\end{table}

\section{Conclusion}
In this work, we described \methodname -- a candidate generation system that was designed as part of our ads multistage ranking system. We hypothesized that a heterogeneous candidate generation system can improve performance without substantially increasing computational costs. We demonstrated that an ensemble of different techniques, each with their own bias, are able to provide significant improvements in both offline metrics and product metrics through the use of online A/B tests. As part of these experiments, we also outline a suite of metrics monitor various aspects of the complete ads ecosystem. Finally, we outline multiple exciting new directions to explore to cost-effectively expand this heterogeneous strategy.

\bibliographystyle{unsrt}
\bibliography{bibliography}

\end{document}